\documentclass[conference]{IEEEtran}
\IEEEoverridecommandlockouts
\usepackage{cite}
\usepackage{amsmath,amssymb,amsfonts}
\usepackage{algorithm}
\usepackage{algorithmic}
\usepackage{graphicx}
\usepackage{subfigure}
\usepackage{textcomp}
\usepackage[small]{caption}
\usepackage{booktabs}
\usepackage{multirow}
\usepackage{xcolor}
\usepackage{float}
\usepackage{bm}
\def\BibTeX{{\rm B\kern-.05em{\sc i\kern-.025em b}\kern-.08em
    T\kern-.1667em\lower.7ex\hbox{E}\kern-.125emX}}
\begin{document}

\title{Efficient Spatial Nearest Neighbor Queries Based on Multi-layer  Voronoi Diagrams\\
\thanks{This  work  is  supported  by  the  National  Natural  Science Foundation of China under Grant U1711267.}
}
\author{
\IEEEauthorblockN{Yang Li}
\IEEEauthorblockA{\textit{School of Computer Science} \\
\textit{China University of Geosciences}\\
Wuhan, China \\
liyang\_cs@cug.edu.cn}
\and
\IEEEauthorblockN{Gang Liu*}
\IEEEauthorblockA{\textit{School of Computer Science} \\
\textit{China University of Geosciences}\\
Wuhan, China \\
liugang67@163.com}
\and
\IEEEauthorblockN{Junbin Gao}
\IEEEauthorblockA{\textit{The University of Sydney Business School} \\
\textit{The University of Sydney}\\
Sydney, NSW, Australia \\
junbin.gao@sydney.edu.au}
\and
\IEEEauthorblockN{Zhenwen He}
\IEEEauthorblockA{\textit{School of Computer Science} \\
\textit{China University of Geosciences}\\
Wuhan, China \\
zwhe@cug.edu.cn}
\and
\IEEEauthorblockN{Mingyuan Bai}
\IEEEauthorblockA{\textit{The University of Sydney Business School} \\
\textit{The University of Sydney}\\
Sydney, NSW, Australia \\
mbai8854@uni.sydney.edu.au}
\and
\IEEEauthorblockN{Chengjun Li}
\IEEEauthorblockA{\textit{School of Computer Science} \\
\textit{\;China University of Geosciences\;}\\
Wuhan, China \\
cuglicj@126.com}
}

\maketitle

\begin{abstract}
Nearest neighbor (NN) problem is an important scientific problem.
The NN query, to find  the closest one to a given query point among a set of points, is widely used in applications such as density estimation, pattern classification, information retrieval and spatial analysis. 
A direct generalization of the NN  query is the $k$ nearest neighbors ($k$NN) query, where the $k$ closest point are required to be found. 
Since NN and $k$NN problems were raised, many algorithms have been proposed to solve them.
It has been indicated in literature that the only method to solve these problems exactly with sublinear time complexity, is to filter out the unnecessary spatial computation by using the pre-processing structure, commonly referred to as the spatial index.

The recently proposed spatial indexing structures which can be utilized to NN search are almost constructed through spatial partition. These indices are tree-like, and the tree-like hierarchical structure can usually significantly improve the efficiency of NN search.
However, when the data are distributed extremely unevenly, it is difficult to satisfy both the balance of the tree and the non-overlap of the subspace corresponding to the nodes. Thus the acceleration performance of the tree-like indices is severely jeopardized.   

In this paper, we propose a non-tree spatial index which consists of multiple layers of Voronoi diagrams (MVD). This index structure can entirely avoid the dilemma  tree-like structures face, and solve the NN problems stably with logarithmic time complexity. Furthermore, it is convenient to achieve $k$NN search by extending NN search on MVD. In the experiments, we evaluate the efficiency of this indexing for both NN search and $k$NN search by comparing with VoR-tree, R-tree and kd-tree. The experiments indicate that compared to NN search and $k$NN search with the other three indices, these two search methods have significantly higher efficiency with MVD.
\end{abstract}


\section{Introduction}

\nocite{*}Nearest neighbor (NN) is a very important problem in the field of information science and data science\cite{DBLP:books/aw/Knuth73}.
NN query, to find the closest point among a set of points to a given query point, is widely required in several applications such as density estimation, pattern classification, information retrieval and spatial analysis.
A direct generalization of NN query is $k$ nearest neighbors ($k$NN) query, where we need to find the $k$ closest points.
According to the representation of space and the measurement of distance, there are many variations of NN/$k$NN query.
However, most problems related to NN/$k$NN query in the real world can be reasonably described based on Euclidean space.
Hence, in this paper, we study the NN and $k$NN problem in Euclidean space, where the distance between points is measured by Euclidean distance.

Given two points $A=\{a_1, a_2,...,a_d\}$ and $B=\{b_1, b_2,...,b_d\}$ in $\mathbb{R}^d$, the Euclidean distance between $A$ and $B$, $\parallel A-B \parallel$, is defined as follows:
\begin{align}
\parallel A-B \parallel=\sqrt{\sum_{i=1}^d(a_i-b_i)^2}
    \label{eq1}
\end{align}
Mathematically, the NN/$k$NN in Euclidean space can be stated as follows.
Given a set $P$ of points in $\mathbb{R}^d$ and a query point $q \in \mathbb{R}^d$, a point $p'$ can be called NN($P$, $q$), the nearest neighbor of $q$ in $P$, if and only if it satisfies the following condition:
\begin{align}
\forall p \in P,\; \parallel p'-q\parallel \leq \parallel p-q\parallel
    \label{eq2}
\end{align}
Similarly, given a set $P$ of points in $\mathbb{R}^d$ and a query point $q \in \mathbb{R}^d$, a set $P'$ of $k$ points can be called $k$NN($P$, $q$), the $k$ nearest neighbors of $q$ in $P$, if and only if it satisfies the following condition:
\begin{align}
\forall p' \in P',\; \forall p \in P\setminus P',\; \parallel p'-q\parallel \leq \parallel p-q\parallel
    \label{eq3}
\end{align}

Theoretically, it is not difficult to find the nearest point to a test point $q$ among a point set $P$. 
The most straightforward way for the purpose is to compute the distance between $q$ and each point in $P$ and then find the point with minimum distance.
Since the query time of the above full search method is proportional to the size of the point set, it is often called linear time algorithm.
In practice, however, this approach is not commonly used because of its intolerable inefficiency when faced with large-scale problems.
In consequence, how to solve it with a sub-linear complexity is the key to this problem.

As the saying goes, there is no such thing as a free lunch. It is impossible to determine the nearest neighbor exactly without calculating the distance between each point in the query and the test point.
Therefore, the only way to reduce the query time of NN is to pre-process the data set into a particular structure and then search from this structure rather than directly from the original data set.
The structures mentioned above to speed up queries are commonly called indices, while those for multidimensional scenarios are referred to as spatial indices.
So far, there have been a lot of research on NN/$k$NN problem, and many spatial indices for this problem have been proposed successively.The majority of the existing spatial search methods are tree-structured. Their inventors hope to realize spatial query in logarithmic computational complexity by taking advantage of the multi-level feature of tree structure. 

In order to realize the multi-level structure, kd-tree adopts a spatial dichotomy strategy. The problem space is divided into a large number of subspaces and distributed to the nodes of the tree. Generally, it performs well on the NN query. However, when the spatial distribution of data is very uneven, the tree structure is easily unbalanced, that is to say, the depth of partial sub tree is much larger than that of other sub trees. Therefore, the NN query in this case is difficult to achieve the theoretical logarithmic computational complexity. 

R-tree uses rectangles to divided space and builds its tree structure. It uses the node segmentation strategy to achieve the balance of the tree. However, there is often a lot of overlap between nodes. This phenomenon becomes more serious when dealing with unevenly distributed data sets. As a result, R-tree often needs to visit too many leaf nodes to get accurate results in NN query. Although some studies have reduced the overlap between nodes through some strategies, theoretically a dynamic R-tree is difficult to completely eliminate the overlap.

Later, a composite index structure, VoR-tree, was proposed. It integrates a Voronoi diagram into R-tree to improve the $k$NN query of R-tree.
% 现已提出的绝大多数的空间索引都是树形结构的，他们的发明者都希望利用树形结构的多级特性来实现对数级时间复杂度的空间查询。为了实现这种树形的多级结构，kd-tree采用的是一种空间二分策略，将问题的空间划分成大量子空间，分配到树的各级节点上。通常情况下它在最近邻查询上的表现良好，但是在数据的空间分布很不均匀的时候，该索引的树形结构很容易失衡，也就是说部分子树的深度远大于其他子树，那么此时的最近邻查询很难实现理论上的对数级计算复杂度。R-tree的做法则是利用矩形来划分空间构建树形结构。它利用节点分割策略实现了树的平衡。但是其节点间经常存在大量的重叠，尤其在面对空间分布不均的数据集时，这一现象会更加严重。所以R-tree在做最近邻查询的时候经常需要访问过多的叶子节点才可以准确的得到结果。虽然有一些研究通过某些策略减了节点之间的重叠，但是理论上，一个动态的R-tree很难完全消除节点间的重叠。后来另一种组合结构的索引结构VoR-tree被提出，这种索引结构将一个维诺图融入R树中，来改善R-tree的$k$NN查询。
Efficient processing of nearest neighbor queries requires spatial data structures which capitalize on the proximity of the objects to focus the search of potential neighbors only\cite{Sharifzadeh:2010:VRV:1920841.1920994}.
% 所以VoR-tree利用R-tree的树形结构来实现最近邻查询，然后利用维诺图中点与点之间的近邻关系，实现一系列其他的近邻相关的查询。相比R-tree，VoR-tree在$k$NN, R$k$NN, kANN和SSQ集中查询上的性能都得到了有效改善。
Therefore, the VoR-tree utilizes the tree structure to achieve the nearest neighbor search, and then realize a series of the other nearest neighbor related search through the pointwise nearest neighbor relations in the Voronoi diagram. Compared with R-tree, VoR-tree has an effective enhancement of the performance on $k$NN, R$k$NN, $k$ANN and SSQ integrated search.

\begin{figure}[htbp]
    \centering
    \includegraphics[width=0.35\textwidth]{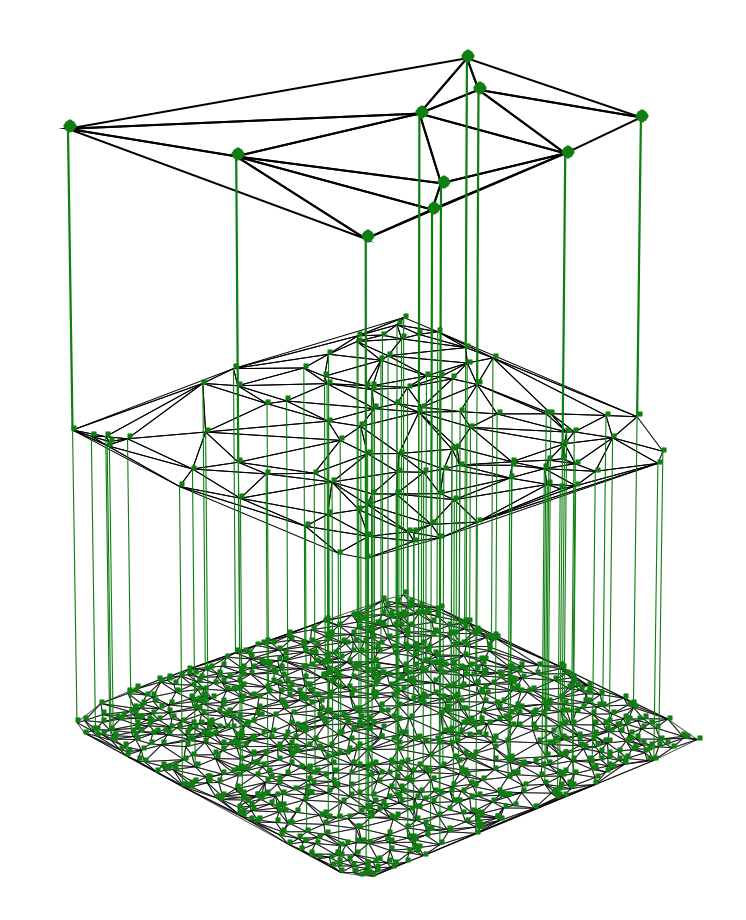}
    \caption{The structure of MVD}
    \label{P_MVD}
\end{figure}
However, since its main structure is R-tree, its efficiency of NN query is almost the same as that of R-tree. In order to completely avoid the disadvantages of tree-structured index, we propose a non-tree spatial index, MVD, which consists of multiple layers of Voromoi diagrams, as shown in Figure~\ref{P_MVD}. Since there is neither branching structure of the tree nor spatial partition, this structure does not have the imbalance problem like the tree structure, and there is no node overlap problem. We also propose methods of batch creation and update maintenance of MVD, thus this index is also applicable to dynamic data. We evaluate the performance of MVD on NN and $k$NN through comparison experiments with kd-tree, R-tree and VoR-tree. The experiments indicate that the performance of our index is better than the other three indexes both with virtual data and real world data, when the data dimension is no more than 4.

\section{Related works}
\subsection{kd-tree}
The kd-tree\cite{Bentley:1975:MBS:361002.361007}, a special case of binary space partitioning tree, is a spatial data structure for organizing points in a k-dimensional space. 
It is probably the simplest data structure available for nearest neighbor searches.

The average time complexity of the NN search based on kd-tree is $O(log\,n)$\cite{Friedman:1977:AFB:355744.355745,DBLP:journals/acta/LeeW77}. However, after inserting a large number of new points, kd-tree tends to lose its balance and the efficiency of the NN search based on it decreases as well in consequence. Some past studies \cite{Procopiuc2003Bkd,DBLP:conf/sigmod/Robinson81,DBLP:conf/dimacs/Maneewongvatana99} attempt to improve the performance of kd-tree from the perspective of spatial partition or the reconstruction of balance. It is effective to some extent, but it is still very difficult to enhance its spatial search performance when facing the extremely unevenly distributed data.
% 基于kd-tree的NN查询的平均时间复杂度是$O(log\,n)$\cite{Friedman:1977:AFB:355744.355745,DBLP:journals/acta/LeeW77}。但是经过大量的新点插入之后，kd-tree会逐渐失去平衡，基于它的NN查询效率也会严重下降。有一些研究\cite{Procopiuc2003Bkd,DBLP:conf/sigmod/Robinson81,DBLP:conf/dimacs/Maneewongvatana99}试图从空间划分方式上或平衡重建上去改善kd-tree的性能，在一定程度上是有效果的。但是在面对非常极端的不均匀分布数据，其空间查询的性能依然很难提高。
\subsection{R-tree}
The R-tree\cite{Guttman:1984:RDI:602259.602266}, proposed by Antonin Guttman in 1984, is perhaps the most widely used spatial index structure. R-tree is considered to be a generalization of B-tree\cite{DBLP:conf/sigmod/BayerM70} in multidimensional space. Contrast with kd-tree, R-tree can organize not only points, but also spatial objects of none-zero size. The key idea of R-tree is to recursively group nearby objects in $\mathbb{R}^d$ by using $d$ dimensional minimum bounding rectangles (MBR). Similar to B-tree, R-tree  maintains its balance by splitting overflow nodes. 

The Depth-First (DF) algorithm \cite{DBLP:conf/sigmod/RoussopoulosKV95} realizes the NN search on the R-tree. The DF algorithm applies Mindist and MinMaxdist to prune the R-tree pruning, followed by using the search distance of the recursive deep-first strategy to search the nearest point. Then the DF algorithm is improved by \cite{Cheung:1998:ENN:290593.290596}. In this improved method, the Mindist is the condition to utilize the deep-first to traverse the R-tree. A Best-First (BF) algorithm \cite{DBLP:journals/tods/HjaltasonS99} is also developed. Its main characteristic is to use the priority queue to save the visited nodes by the Mindist rank of nodes.  
It is the state-of-the-art NN/$k$NN algorithm of R-tree.
Because R-tree is balanced, it tend not to be too deep. However, in this structure, the rectangles often overlap so much that more subtrees need to be searched during spatial queries. Therefore，R-trees generally perform well, but do not guarantee good worst-case performance\cite{Performance_Evaluation_of_Main-Memory_R_tree_Variants}.

For the optimization of the R-tree structure, most of scholars choose to alternate the node separation strategy to deduce the overlap between nodes \cite{Ang:1997:NLN:647225.718938,DBLP:conf/icde/Greene89}. Among their proposed methods, the most effective refinement might be R$^*$-tree\cite{Beckmann:1990:RER:93597.98741}, whereas the priority R-tree \cite{Arge:2004:PRP:1007568.1007608} enhances the worst-case performance of R-tree.

\subsection{VoR-tree}
The VoR-tree\cite{Sharifzadeh:2010:VRV:1920841.1920994} is a state-of-the-art R-tree based spatial index. A VoR-tree is a combination of an R-tree and a Voronoi diagram. This index structure benefits from both the neighborhood exploration capability of Voronoi diagrams and the hierarchical structure of R-tree. 

\section{Background}
%%%%%% 维诺图
\subsection{Voronoi diagram and its properties}

The Voronoi diagram\cite{DBLP:books/wi/OkabeBSCK00}, proposed by Rene Descartes in 1644, is a spatial partition structure widely applied in many science domains, most notably spatial database and computational geometry. In a Voronoi diagram of $n$ points, the space is divided into $n$ regions corresponding to these points, which are called Voronoi cells.
For each point of these $n$ points, the corresponding Voronoi cell consists of all locations closer to that point than to any other.
In other words, each point is the NN of all the locations in its corresponding Voronoi cell.
Formally, the above description can be stated as follows.
Given a set $P$ of $n$ points, the Voronoi cell of a point $p \in P$, written as $V(P, p)$, is defined as \eqref{eq4}
\begin{align}
V(P, p)=\{q\;|\;
\forall p' \in P,\; p' \neq p,\;\parallel p-q\parallel \leq \parallel p'-q\parallel\}
    \label{eq4}
\end{align}
and the Voronoi diagram of $P$, written as $VD(P)$, is defined as \eqref{eq5}.
\begin{align}
VD(P)=\{V(P, p)\;|\; p \in P\}
    \label{eq5}
\end{align}
The Voronoi diagram has the following properties:\\
\textbf{Property\,1}: The Voronoi diagram of a certain set $P$ of points, $VD(P$), is unique.\\
\textbf{Property\,2}: Given the Voronoi diagram of $P$, the nearest point of $P$ to a point $q\in P$ is among the Voronoi neighbors of $q$. That is, the closest point to $q$ is one of generator points whose Voronoi cells share a Voronoi edge with $V(P,q)$. \\
\textbf{Property\,3}: Given the Voronoi diagram of $P$ and a test point $q \notin P$, a point $p'$ is the nearest point of $P$ to $p$, if and only if $q \in V(P,p')$.\\
\textbf{Property\,4}: Property 2 and Property 3 suggests that, given a point set $P$ and a test point $q \notin P$, the second nearest point of $P$ to $q$ is among the Voronoi neighbors of $NN(P,q)$.\\
\textbf{Property\,5}: By generalizing Property\,4, we can obtain that, Let $p_1,\,p_2,\,\cdots,\, p_k$ be the $k>1$ nearest points
of $P$ to a query point $q$ (i.e., $p_i$ is the $i$-th nearest neighbor of
q), then, $p_k$ is a Voronoi neighbor of at least one point
$p_i \in  \{p_1,\,p_2,\,\cdots,\, p_{k-1}\}$ ($p_k \in VN(p_i)$)\cite{Sharifzadeh:2010:VRV:1920841.1920994}.\\
\textbf{Property\,6}: Let $n$, $n_e$ and $n_v$ be the number of generator points, Voronoi edges and Voronoi vertices of a Voronoi diagram in $\mathbb{R}^2$, respectively, and assume $n \geq 3$. Then,
\begin{align}
n+n_v-n_e=1
    \label{eq6}
\end{align}
Every Voronoi vertex has at least 3 Voronoi edges and each Voronoi edge  belongs to two Voronoi vertices, hence the number of Voronoi edges is not less than $3(n_v+1)/2$, i.e.,
\begin{align}
n_e \geq \frac{3}{2}(n_v+1)
    \label{eq7}
\end{align}
According Formula \eqref{eq6} and Formula \eqref{eq7},the following relationships holds:
\begin{align}
n_e \leq 3n-6
    \label{eq8}
\end{align}
\begin{align}
n_v \leq 2n-5
    \label{eq9}
\end{align}
\textbf{Property\,7}: When the number of generator points is large enough, the average number of Voronoi  edges  per  Voronoi  cell  of a Voronoi diagram in $\mathbb{R}^d$  is a constant value depending only on $d$. When $d$ = 2, every Voronoi edge is shared by two Voronoi Cells. Hence, the average number of Voronoi edges per Voronoi cell dos not exceed 6, i.e., $2 \cdot n_e/n =2(3n-6)/n= 6-12/n \leq 6$.

\subsection{Delaunay triangulation and its properties}
\begin{figure}[t]
    \centering
    \includegraphics[width=0.49\textwidth]{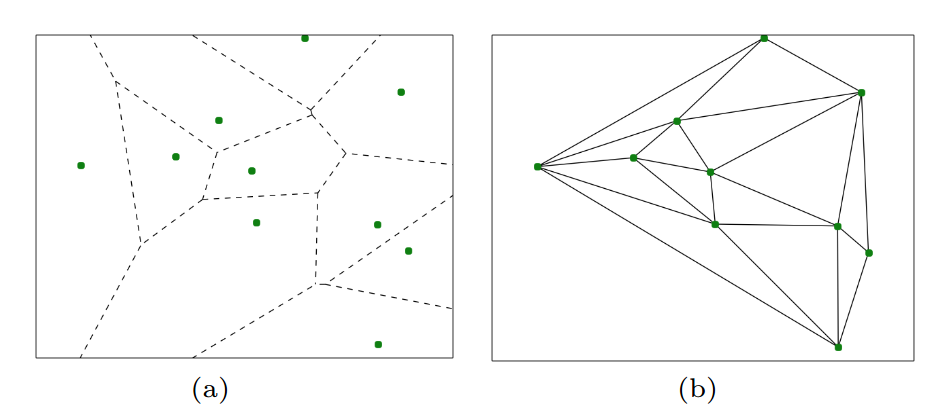}
    \caption{a) Voronoi Diagram, b) Delaunay Graph}
    \label{P_VD}
\end{figure}
Delaunay triangulation\cite{Delaunay1934Sur} is a very famous triangulation proposed by Boris  Delaunay in 1934. For a set $P$  of discrete points in a plane, the Delaunay triangulation $DT(P)$ is such a triangulation that no point in $P$ is inside the circumcircle of any triangle of $DT(P)$. 
The Delaunay  triangulation has the following properties:
\\
\textbf{Property\,8}: The Delaunay triangulation of a set of points is dual to its Voronoi diagram.\\
\textbf{Property\,9}: A graph of Delaunay triangulation must be a connected graph, that is, any two vertices in the graph are connected.\\
\textbf{Property\,10}: For a set of points, its nearest neighbor graph is a subgraph of its Delaunay triangulation graph.\\
\textbf{Property\,11}: The Delaunay triangulation of $n$ points in $\mathbb{R}^d$ contains $O(n^{d/2}$) simplices, where a $d$ dimensional simplex, the generalization of triangle, consists of $d+1$ points.\\
\section{Structure of MVD}
In this section, we introduce the structure of MVD, a novel non-tree spatial index we proposed, and describe its construction method in detail.

\begin{figure}[htbp]
  \centering
  \subfigure[]{\includegraphics[width=0.4\textwidth]{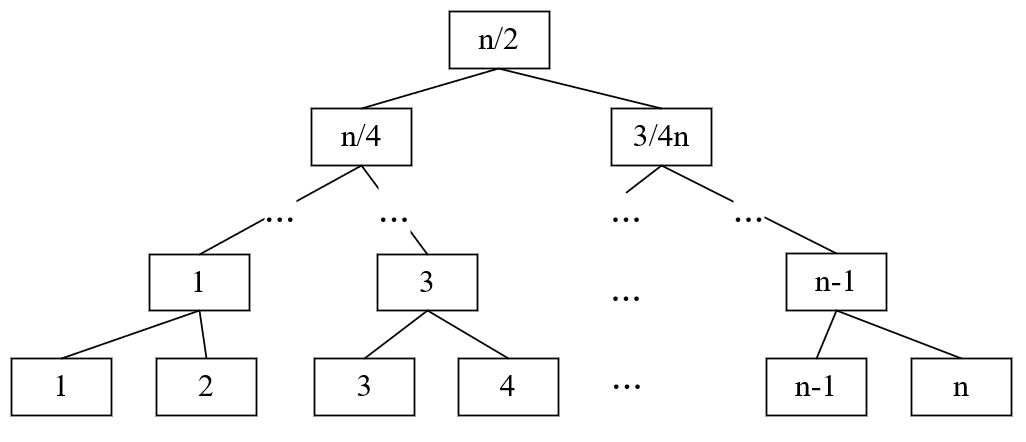}}
  \subfigure[]{\includegraphics[width=0.4\textwidth]{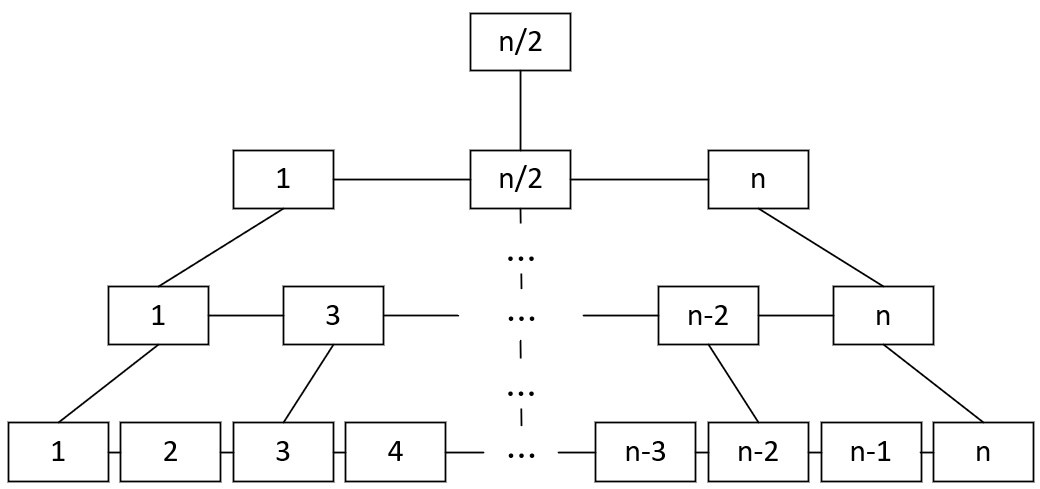}}
  \caption{a) Binary sort tree, b) Multilayer linked list }
  \label{structure_compare}
\end{figure}

As aforementioned, all the tree spatial indexing are through distributing the data points to the Multi-layer  subspace. This process can achieve the NN search with logarithmic time complexity. On this indexing structure, it is difficult to satisfy both the balance of the tree and the independence of the subspace corresponding to the nodes. However, the unbalance of the tree and the spatial overlap of the nodes severely affect the efficiency of the spatial search.  In the past literature, the neighboring relations in the physical space are projected to the network topology structure and this network structure helps to accelerate the NN related spatial search. The VoR-tree is one of the typical examples. This combination of indexing strategy can effectively improve the $k$NN search to a certain extent, but the cause of its logarithm search efficiency of $NN$ search is still its tree main indexing structure. Thus, it cannot completely solve the weaknesses of the tree indexing structures.
Hence to avoid this dilemma entirely, it is only possible to abandon the tree structure to design a new non-tree indexing structure.

%树形结构的索引之所以在很多情况下可以带来对数级的检索效率，只因为它具备一种金字塔形状，也就是从顶层根节点开始，逐级向下，数据的规模呈指数增长。如图2（a）展示了一个管理着n个数字的二分搜索树。我们根据这个结构可以轻易的经过O(log\,n)次访问找到其中任意一个数查找到其中任意一个数字。
The tree structure indexing can always bring the logarithm search efficiency, because it has a pyramid structure, namely, from the top root node travelling down layer by layer with a logarithm growth of the data scale. As in Figure \ref{structure_compare}(a), we demonstrate a binary search tree which manages $n$ numbers. Based on this structure, we can easily find any number by using $O(log\,n)$ visits to obtain any other number.

%只有将数据按一定规则逐级划分，才可以实现这种对数级复杂度的检索吗？答案当然是否定的。如图2（b）展示了一个3层链表链表结构，每两层都间都有若干个链接。该结构的底层有序的存储着1到n这n个数字，第二层存储的则是从第一层中按照1/2的等间隔抽取上来的数字，显然这个链表也是有序的。按照同样的规则逐层构建，便形成了这样一种结构。通过这样一种一种结构，我们同样可以只经过O(log\,n)次点的遍历找其中任意一个点。这样一种结构只需要保证每层链表是有序的，即次序上相邻的节点之间都是相连的，便可以实现对数级的复杂度的查询。
Is it only possible to achieve the logarithm complexity search by the layer-by-layer partition of the data with a certain rules? The answer is certainly no. Figure \ref{structure_compare}(b) demonstrates a 3-layer list structure where there are several connections between every two layers. At the bottom layer of this structure, there are $n$ integer values from $1$ to $n$ stored in order. The second layer stores the numbers sampled from the first layer with the interval of $1/2$. Evidently, this list is also ordered. Constructing the structure with the same rule, we can have this specific structure. From this specific structure, we can also find any point by traversal with $O(log\,n)$ times. Therefore, as long as we guarantee the list on each layer is ordered, i.e., the nearest nodes are connected in terms of the order, the search can be accomplished with the logarithm complexity.

%如果我们将该结构推广到2维或者更高维度，我们可以用构建维诺图的来代替排序操作，将空间上相邻的点连接起来。然后按一定的比例从中均匀的抽取数据点继续构建维诺图，作为上一层。不断重复上述操作，直至当前层的数据量已经足够小。我们将这种由多层维诺图结构的结构命名为MVD。上述创建MVD的过程的伪代码如\ref{alg:MVD}所示，其中参数k被称之为构建系数，表示表示相邻层之间，下层数据量与上层数据量的k倍。基于MVD的空间最近邻查询方法，我们将在下一节详细介绍。
If we extend this structure to two or higher dimensions, we can replace the sort operation with constructing the Voronoi diagram. The nearest nodes on the space can be connected. Then the data are evenly sampled from them by a certain proportion to be the upper layer. We repeat this process, until the data size of the current layer is small enough. This resulting %structure from the
Multi-layer  Voronoi diagram structure is named as MVD. The pseudo code of the aforementioned process to create the MVD is presented in Algorithm~\ref{alg:MVD}, where the parameter $k$ is referred to as the construction parameter which represents the data size in the lower layer is $k$ times of what is in the upper layer across the layers. For the methods to search for the nearest neighbors based on the MVD space, we are intended to detail them in the next section.

%%%%%%%%% MVD
\begin{algorithm}[ht]
\caption{MVD}
\label{alg:MVD}
\textbf{Input}: The point set $P$\\
\textbf{Parameter}: $k$\\
\textbf{Output}: The Multi-layer  Voronoi Diagrams of $P$, $M_P$

\begin{algorithmic}[1]
\STATE $V :=$ VD($P$);
\STATE $M_P:=$ [$V$];
\WHILE{Size($P$) $> k$}
\STATE $P :=$ Sample($P$, Size($P$)$/k$);
\STATE $V :=$ VD($P$);
\STATE Append($M_P$, $V$);
\ENDWHILE
\STATE \textbf{return} $M_P$.
\end{algorithmic}
\end{algorithm}

\section{Query Processing on MVD}
In this section, we introduce the query processing on MVD from the aspects of NN and $k$NN, and elaborate on their operation mechanism and theoretical time complexity.
\subsection{Nearest Neighbor Query}

Given a Voronoi graph, if we assume that from any generator point we can arrive at its Voronoi neighbor by moving one step only, for any two generator points, at least one connection path exists through several generator points in the neighborhood
based on \textit{Property 8} and \textit{9}. This path is named as the Voronoi connection path. Certainly, there is not only one Voronoi connection path between any two nodes. If we can find one shortest path among those, namely the path through the least number of generators, we can start from any node, traverse the least intermediate nodes and visit the target node.

According to the definition of NN and \textit{Property 3}, we can draw such an inference: given  the Voronoi diagram of a point set $P$ and a test point $q \notin P$, a point $p'$ is the closest point of $P$ to $q$, if and only if it satisfies the following condition:
\begin{align}
\forall p \in VN(P,p'),\; \parallel p'-q\parallel \leq \parallel p-q\parallel
    \label{eq10}
\end{align}
Conversely, we can further make the following corollary by reductio ad absurdum:
given an arbitrary point $p* \in P$ that is not the closest point to $q$, there must exist a point in Voronoi neighbors of p* which is closer to $q$ than $p*$. Formally, it can be written as follows:
\begin{align}
\forall p*& \in P \setminus \{NN(P,q)\},\; \exists p \in VN(P,p*):\nonumber \\ &\parallel p-q\parallel \leq \parallel p*-q\parallel
    \label{eq11}
\end{align}

Based on this deduction, we develop the VD-NN search algorithm. It is an NN search algorithm based on the Voronoi diagram which finds the nearest neighbor node of the target node through the shortest Voronoi connection path. The pseudo code is presented as in Algorithm~\ref{alg:VD_NN}.

Firstly, we heuristically start from any point in the point set $P$. If the Voronoi neighborhood of the point $p$ contains the point that is closer to the $P$ distance search point $q$, then we will move to a point closest to $q$ in $p$'s Voronoi neighborhood. We repeat this process until the point at the current location does not have any point which is closer to point $q$. The point where we terminate at, is the nearest neighbor point of $q$ in the point set $P$.

\begin{algorithm}[ht]
\caption{VD-NN}
\label{alg:VD_NN}
\textbf{Input}: The Voronoi Diagrams $V_P$, the query point $q$ and starting point $p_s$\\
\textbf{Output}: The nearest neighbor point of $q$ $nn_q$

\begin{algorithmic}[1] %[1] enables line numbers
\IF{$p_s \neq $ Null}
\STATE $nn_q := p_s$;
\ELSE 
\STATE $nn_q :=$ Sample($P$);
\ENDIF
\STATE $Visited :=$ \{$nn_q$\};
\STATE $found :=$ False;
\WHILE{\NOT $found$}
\STATE $found :=$ True;
\FORALL{$n \in$ $V_P$[ $nn_q$]}
\IF{$n \notin Visited$}
\STATE $Visited:=Visited \cup \{n\}$;
\IF{$\parallel q-n\parallel < \parallel q-nn_q\parallel$}
\STATE $nn_q := n$;
\STATE $found :=$ False;
\ENDIF
\ENDIF
\ENDFOR
\ENDWHILE
\STATE \textbf{return} $nn_q$.
\end{algorithmic}
\end{algorithm}

For a set $P \in \mathbb{R}^d$ consisting of $n$ points, the mean time complexity of VD-NN is $O(n^{1/d})$. Since the initial point is randomly selected, this value is really uncertain. In the ideal circumstance, when starting from a very close point to the search point, the time complexity becomes extremely low and even can reach the constants. Therefore, the efficiency of VD-NN is very sensitive to the choice of the staring point.
% 面对一个由n个点构成的集合P（in $\mathbb{R}^d$ ），VD-NN的平均时间复杂度是$O(n^{1/d})$，因为起始点的选择是随机的，所以这个值很不稳定。在理想的情况下，从一个距离查询点距离非常近的点出发，那么这个时间复杂度便会变得非常低，甚至可以达到常数级。所以VD-NN的效率对于起始点的选择是非常敏感。

With the MVD structure, we realize the NN search algorithm MVD-NN with logarithm time complexity, through multiple efficient executions of calls for VD-NN.
%利用MVD结构，我们通过多次高效的VD-NN调用实现了一种对数级NN查询算法（MVD-NN）。
Given a $h$-layer MVD generated by the point set $P$ consisting of $n$ points, the following procedure is to search for the nearest neighbor point of $q$ from the point set $P$ applying the MVD-NN algorithm.
% 给定一个由点集P（由n个点构成）生成的一个h层MVD,运用MVD-NN算法从点集合P中查找q的最近邻点的过程是这样的：
Firstly, the starting point is randomly selected from the top of MVD. In subsequence, VD-NN is called to find the point $p_q$ which is closest to $q$. Then we use $p_1$ as the new starting point and call VD-NN to obtain the point $p_2$ closest to $q$. This process is executed recursively. Eventually, we can reach the last layer to attain $p_h$. $p_h$ is the nearest neighbor point of $q$ in $P$. 
% 首先从MVD的顶层中随机选择一个点作为起始点，调用VD-NN找该层中距离q最近的一个点$p_1$,然后将$p_1$作为起始点，调用VD-NN从第2层中找到距离q最近的点$p_2$，重复这个过程从第3层找到$p_3$,从第4层找到$p_4$,...最后从底层找到$p_h$，$p_h$就是q在P中的最近邻点。
When building MVD with $k=10$ and $2$ dimensions, it only needs approximately $\sqrt{10}\approx 3$ nodes to find $p_1$ from a random starting point in the first layer. The reason is that in the second layer, $p_1$ is really close to $p_2$. Hence $p_2$ can also be acquired after 3 nodes starting from $p_1$. Similarly, this procedure appears in the following layers. In each layer of MVD, the time complexity to execute VD-NN is $O(k^{1/d})$ which is actually a constant as $k$ is a constant. However, the number of layers of MVD is $log_kn$. Therefore, the time complexity of MVD-NN is approximately $O(log\, n)$.
%如果构建MVD时，k的值是10，在2维的情况下，从第一层的一个随机点出发只需要经过大约\sqrt{10}=3个节点便可以找到$p_1$，因为在第二层中，$p_1$已经非常接近于$p_2$,所以从$p_1出发经过大约3个节点便可以找到$p_2$，后面的每层同理。在MVD的每一层VD-NN的执行的时间复杂度实是$O(k^{1/d})$(实际上是一个常数，因为k是一个常数)，而MVD的层数是$log_kn$，所以MVD-NN的时间复杂度大约为$O(log\, n)$。

\begin{algorithm}[ht]
\caption{MVD-NN}
\label{alg:MVD_NN}
\textbf{Input}: The Multi-layer  Voronoi Diagrams $M_P$ and the query point $q$\\
\textbf{Output}: The nearest neighbor point of $q$ $nn_q$

\begin{algorithmic}[1] %[1] enables line numbers
\STATE $i:=$ Size($M_P$) $-$  $1$;
\WHILE{$i>0$}
\STATE $V:=M_P$[$i$];
\STATE $nn_q:=$ VD-NN($V$, $q$, $nn_q$);
\STATE $i:=i-1$;
\ENDWHILE
\STATE \textbf{return} $nn_q$.
\end{algorithmic}
\end{algorithm}

\subsection{$k$ Nearest Neighbor Query}
As with VoR-tree, MVD uses the characteristics of the Voronoi graph. It achieves the $k$NN query through the incremental method which finds the second until the $k$-th nearest neighbor point by the extension of the NN search. Nevertheless, our structure still has two major differences with VoR-tree. The first difference is that the $k$NN search of VoR-tree is extended from BFS that is the NN search algorithm based on R-tree, whereas MVD-$k$NN is extended from MVD-NN. Of course, the choice of the basic NN algorithm can indirectly influence the efficiency of $k$NN algorithm. For the second difference, in the realization of the incremental $k$NN algorithm, we use the fixed length ranking array. The reason is discussed in detail in the following description of the algorithm.

% 和VoR-tree相似，MVD也是运用Voronoi图的特性，通过incremental method（也就是通过扩展NN查询，逐一找到第2至第k近的点）来实现的$k$NN查询的。 然而在$k$NN查询的实现上，我们的结构和VoR-tree还是有两点区别的。首先，我们VoR-tree的$k$NN查询（VoR-$k$NN）扩展自BFS(基于R-tree的NN查询算法)，而基于MVD的$k$NN查询（MVD-$k$NN）扩展自MVD-NN；当然，基础NN算法的选择也会间接影响到$k$NN算法的效率。其次，在增量$k$NN算法实现的过程中，VoR-tree选用小根堆来作为备选集合，我们选用是一个定长的排序数组；我们这么做的具体原因将在接下来的算法描述中具体说明。

\begin{figure}[htbp]
    \centering
    \includegraphics[width=0.25\textwidth]{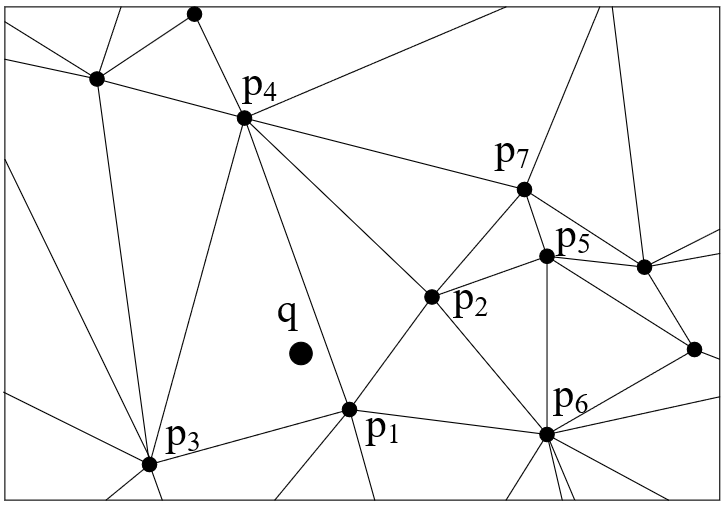}
    \caption{$k$NN of $q$}
    \label{P_TIN}
\end{figure}

%Fig.\ref{P_TIN}展示了一个点集合P的Delaunay graph（表达了P中点之间的维诺相邻关系）和一个点查询点q。如果把用p_i来表示距离q第i近的点（其中i > 1），那么我们将Property 5形式化的表达为这样:
Figure~\ref{P_TIN} demonstrates the Delaunay graph a point set $P$, and a query point $q$. If we use $p_i$ to represent the $i$-th $(i>1)$ closest point to $q$, we can present Property 5 as the following formulae.
\begin{align}
p_i \in \bigcup_{j=1}^{i-1}VN(P,p_j)
    \label{eq12}
\end{align}
% 因为p_i一定不会出现在由q的前i-1个最近邻点构成的集合中，但是这个集合一定是 Formula.ref{eq12}中p_i候选集的一个子集，所以我们可以进一步缩小p_i的候选集，如Formula.ref{eq13}所示。
Since $p_i$ is certainly not possible to appear in the set of the $(i-1)$-th nearest neighbors of $q$ and this set has to be a subset of the candidate set of Formula~\eqref{eq12}, we can further narrow down the candidate set of $p_i$ as shown in Formula~\eqref{eq13}.
\begin{align}
p_i \in \bigcup_{j=1}^{i-1}VN(P,p_j) \setminus \{p_1, p_2,\cdots, p_{i-1}\}
    \label{eq13}
\end{align}
%所以从第1近邻的点开始，通过添加其维诺邻居来更新候选集，紧接着从候选集中找出第2近邻的点，再添加第2近邻点的维诺邻居来更新候选集，然后从候选集中找到第3近邻的点，重复上述过程，直至找到第k近邻的点。这样基于Voronoi的增量$k$NN算法就得以实现。
Therefore, from the first nearest neighbor, by adding its Voronoi neighbor, we update the candidate set. Following that, we can find the second nearest neighbor and include the second nearest neighbor to update the candidate set. The we find the third nearest neighbor from the candidate set. We repeat this process until we obtain the $k$-th nearest neighbor. From this whole process, we can realize the incremental $k$NN algorithm based on Voronoi.
%如果每次更新候选集和从候选集中查找最近邻的点都可以当做一个原子操作，那么上述算法的时间复杂度就是$O(k)$。为了提升更新候选集和从中查找最近邻点的效率，在VoR-$k$NN算法中，候选集中的每个点都存在一个小根堆（min heap）中，如果候选集大小是m，那么对这个候选集添加（add）和弹出（pop）操作的时间复杂度都是O(log\,m)。由property 7 可以得到在2维的情况下，m \leq 6k。综合来看，基于VoR-tree的$k$NN算法的时间复杂度近似于$O(log\,n+k \cdot log\,k)$。
If the process to update the candidate set and search the nearest neighbor from the candidate set can be seen as an atomic operation, the time complexity of the this algorithm is $O(k)$. In order to improve the efficiency of the update and search process, in the VoR-tree, there is a minimum heap in each point of the candidate set. If the size of the candidate set is $m$, the operation to add and pop to this candidate set is $O(\log\,m)$. From Property 7, in the circumstance to obtain 2 dimensions, $m \leq 6k$. In general, the time complexity of the $k$NN algorithm based on the VoR-tree is approximately $O(log\,n+k \cdot log\,k)$.

%在MVD-$k$NN算法(伪代码如 Algorithm  \ref{alg:MVD_$k$NN}所示)中，我们采用了另外一个长度为k的数组来存储候选集。因为在$k$NN查询中，我们只需要获取查询点的前k个最近邻点，所以，对于一个点，如果在某个阶段中，已经存在k个以上点比它距离查询点更近，那么这个点便可以从候选集中淘汰了。为了使这种淘汰策略效率更高，整个算法的实行过程中我们需要保证候选集数组是有序。整个对于每一个插入新点的操作，最坏的时间复杂度是O(k),最好的时间复杂度是O(1)。所以MVD-$k$NN的时间复杂度介于$O(log\,n+k)$~$O(log\,n+k^2)$之间。然而，很显然若果一点个p比另一个点p‘距离点q更近，那么点某个p的维诺邻居比点p’的某个维诺邻居距离点q的距离更近的概率是非常大。所以在维护上述的这个有序数组时，每次插入的新点在数组中的次序往往非常接近于k，甚至这个点可能会直接被淘汰。这也就意味着在多数情况下MVD-$k$NN的 时间复杂度更接近于$O(log\,n+k)$，当然这并不稳定。但是相比VoR-$k$NN，MVD-$k$NN因为维护了一个更小的临时候选集，所以运行过程中的内存开销是远小于前者的，尤其是在维度较高的时候。
\begin{algorithm}[ht]
\caption{MVD-$k$NN}
\label{alg:MVD_$k$NN}
\textbf{Input}: The Multi-layer  Voronoi Diagrams $M_P$, the query point $q$ and $k$ \\
\textbf{Output}: The $k$NN list $K$
\begin{algorithmic}[1] %[1] enables line numbers
\STATE
$nn_q:=$ MVD-NN($M_P$, $q$);
\STATE 
$K:=$ [$nn_q$];
\STATE 
$Visited:=\{nn_q\}$;
\STATE 
$V_P:=M_P$[$0$];
\FOR{$i:=1$  to $k-1$}
\FORALL{$n \in$ $V_P$[ $K$[$i$]]}
\IF{$n \notin Visited$}
\STATE $Visited:=Visited \cup \{n\}$;
\FOR{$j:=i+1$  to Size($K$)}
\IF{
$\parallel q-n\parallel < \parallel q-K$[$j$] $\parallel$}
\STATE Insert($K$, $j$, $n$);
\IF{Size($K$) $>k$}
\STATE Pop($K$);
\ENDIF
\STATE \textbf{break};
\ENDIF
\ENDFOR
\IF{Size($K$) $<k$ \AND $n  \notin K$}
\STATE Append($K$, $n$);
\ENDIF
\ENDIF
\ENDFOR
\ENDFOR
\STATE \textbf{return} $K$.
\end{algorithmic}
\end{algorithm}

In the MVD-$k$NN algorithm whose pseudo code is presented in Algorithm~\ref{alg:MVD_$k$NN}, we utilize the candidate set with the length of $k$. Since we only need to acquire the $k$-th nearest neighbors before the search point in the $k$NN search, we can eliminate one point from the candidate set if there are more than $k$ points being closer to the search point at a certain stage. With the purpose to improve the efficiency of this elimination strategy, we have to ensure the candidate set to be ordered in the whole execution process of this algorithm. For the whole operation of every inserted new point, the worst-case time complexity is $O(k)$ and the best-case is$O(1)$. Therefore, the time complexity of MVD-$k$NN is between $O(log\,n+k)$ and $O(log\,n+k^2)$. However, it is evident that if a point $p$ is closer to $q$ than the point $p'$, there is an extremely high probability to have a smaller distance between a Voronoi neighbor of $p$ and $q$ than the distance between a Voronoi neighbor of $p'$ and $q$. Thus, when maintaining this ordered array, every time we insert a new point, the order of this new point is usually very close to $k$ and this point even can be eliminated straightforwardly. It means that the time complexity of the MVD-$k$NN can be closer to $O(log\,n+k)$ in most cases. It is certainly not stable, but the MVD-$k$NN maintains a smaller temporary candidate set compared with the VoR-$k$NN and therefore the memory cost of execution is much smaller than the VoR-$k$NN, especially when the dimension is high.

%%%%%%%%% MVD-$k$NN

\section{Maintenance of MVD}
% 我们在前面的章节中已经介绍过如何通过一个给定的点集合创建一个MVD结构。然而，一个时空只有具备包括删除点和添加点在内的动态更新机制，才可以在面对动态数据时仍然保持可用。本节我们将介绍MVD结构的插入算法MVD-Inser（伪代码如 \ref{alg:MVD_INSERT}）和删除算法MVD-Delete（伪代码如 \ref{alg:MVD_DELETE}）
As discussed in the sections above, we introduce how to create an MVD structure based on a given point set. However, a  spatial index can still be used for dynamic data, if it contains the dynamic updating mechanism including deleting and adding the points. In this section, we describe the insert algorithm: MVD-Insert and the deleting algorithm: MVD-Delete with the MVD structure. The pseudo code is demonstrated in Algorithm~\ref{alg:MVD_INSERT} and Algorithm~\ref{alg:MVD_DELETE} respectively.

% 假设存在一个有点集合$P$个构成的MVD索引$M_P$，它的创建系数为$k$。向其插入一个新点$p$的过程是这样的：首先通过Voronoi diagram 的插入算法（VD-Insert）向$M_P$底层的添加$p$（最优秀的VD-Insert 算法平均时间复杂度是$O(log\,n)$）。如果还存在上一层，则将目光移向上一层，继续通过VD-Insert算法以$1/k$的概率添加向$p$这个点，如果该层$p$被插入成功，则继续逐级的以$1/k$的概率插入$p$,直至某一层$p$没有被插入，则整个过程终止。如果顶层也被插入成功，我们将通过则继续以$1/k$的概率向$M_P$添加一层，并向该层中插入$p$。很显然，随着层数的升高，点$p$在当前层被插入的概率也是逐层下降的。这样的一种策略保证了经过大量的新点插入后，$M_P$中的每一层点的数量仍然是上一层的$k$倍。
Assume that there is an MVD indexing $M_P$ which consists of the point set $P$. The process to insert a new point $p$ is as follows. Firstly, we add $p$ to the bottom layer of $M_P$ by the VD-Insert algorithm of the Voronoi diagram. The best VD-Insert algorithm has the mean time complexity as $O(log\,n)$. Secondly, if there is an upper layer, we move one layer up and add $p$ with the probability of $1/k$ by the VD-Inser algorithm. Thirdly, if we successfully insert $p$ in this layer, we still insert $p$ with the same probability step by step. Until $p$ cannot be inserted in a certain layer, we terminate the whole process. If we also succeed in inserting $p$ to the top layer, we can add one more layer to $M_P$ with the probability of $1/k$ and insert $p$. Obviously, as the number of layers increases, the probability of $p$ to be inserted to this layer is decreased. This strategy guarantees that after inserting a large number of new points, the number of the points in each layer of $M_P$ is still $k$ times to the last layer.
% MVD-Insert与一样，当MVD-Dlete执行完之后，$M_P$中，相邻层之间点数量的比例需要保持不变。所以MVD-Dlete的执行过程是这样设计的：遍历$M_P$的每一层，如果每一层，如果某层含有$p$,则利用VD-Delete（Voronoi diagram的删除算法，时间复杂度为$O(1)$）将$p$从这一层中移除，并将$p$在下一层中的最近邻点以$1-1/k$的概率插入该层；否则以$1/k$的概率将$p$从这一层中移除。
Similar with the MVD-Insert, after the execution of the MVD-Delete, the proportion of the number of points between the near layers should be maintained the same in $M_P$. Therefore, the execution of the MVD-Delete is as follows. We traverse all the layer of $M_P$. For every layer, if one layer contains $p$, we use the VD-Delete as the deleting algorithm in the Voronoi diagram with the time complexity $O(1)$ and eliminate $p$. Also we insert the nearest neighbor of $p$ at the lower layer with the probability $(1-1/k)$. Otherwise $p$ is eliminated in this layer with the probability $1/k$.
% 从上述的算法描述中，看一很容易分析出，MVD-Insert和MVD-Dlete的时间复杂度分别是$O(log^2n)$和 $O(log\,n)$。
From the above description of the algorithm, it is obvious that the time complexities of the MVD-Insert and the MVD-Delete are $O(log^2n)$ and $O(log\,n)$ respectively.

%%%%%%%%% MVD-INSERT
\begin{algorithm}[ht]
\caption{MVD-Insert}
\label{alg:MVD_INSERT}
\textbf{Input}: The Multi-layer  Voronoi diagrams $M_P$ and the point $p$ to be inserted\\
\textbf{Parameter}: $k$
\begin{algorithmic}[1]
\STATE $V := M_P$[$0$];
\STATE VD-Insert($V$, $p$);
\FOR{$i:=1$ \TO Size($M_P$)}
\IF{Random($0$, $1$) $<1/k$}
\IF{$i<$ Size($M_P$)}
\STATE VD-Insert($M_P$[$i$], $p$);
\ELSE
\STATE $V :=$ VD($\{p\}$);
\STATE Append($M_P$, $V$);
\STATE \textbf{break};
\ENDIF
\ELSE
\STATE \textbf{break};
\ENDIF
\ENDFOR
\end{algorithmic}
\end{algorithm}

%%%%%%%%% MVD-Delete
\begin{algorithm}[ht]
\caption{MVD-Delete}
\label{alg:MVD_DELETE}
\textbf{Input}: The Multi-layer   Voronoi Diagrams $M_P$ and the point $p$ to be deleted\\
\textbf{Parameter}: $k$

\begin{algorithmic}[1]
\STATE $V := M_P$[$0$];
\STATE VD-Delete($V$, $p$);
\FOR{$i:=1$ \TO Size($M_P$) $-1$}
\STATE $V:=M_P$[$i$];
\IF{$p \in V$}
\STATE VD-Delete($V$, $p$);
\IF{Random($0$, $1$) $<1-1/k$}
\STATE $V':=M_P$[$i-1$];
\STATE VD-Insert($V$, NN($V'$, $p$));
\ENDIF
\ELSIF{Random($0$, $1$) $<1/k$}
\STATE $p:=$ NN($V$, $p$);
\STATE VD-Delete($V$, $p$);
\ENDIF
\IF{Size($V$) $=0$}
\STATE Remove($M_P$, $V$);
\ENDIF
\ENDFOR
\end{algorithmic}
\end{algorithm}

\section{Experiments}
% 在前文中，我们已经从理论上分析了MVD的优越性。在本节中，我们将通过实验来验证MVD在空间最近邻查询上的性能表现。
In the previous content, we discuss the theoretical advantages of the MVD. In this section, we intend to evaluate the performance of the MVD on the spatial nearest neighbor search with numerical experiments.
\subsection{Experimental settings}
\begin{figure}[htbp]
    \centering
    \includegraphics[width=0.52\textwidth]{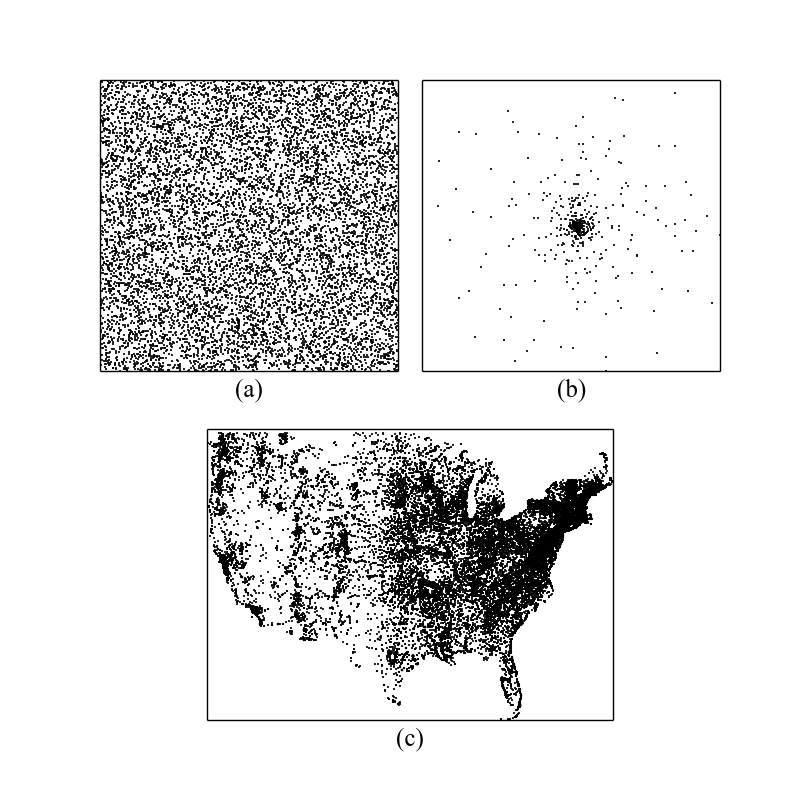}
    \caption{a) Uniform data, b) Nonuniform data, c) US data}
    \label{benchmark}
\end{figure}

\begin{figure}[htbp]
    \centering
    \includegraphics[width=0.48\textwidth]{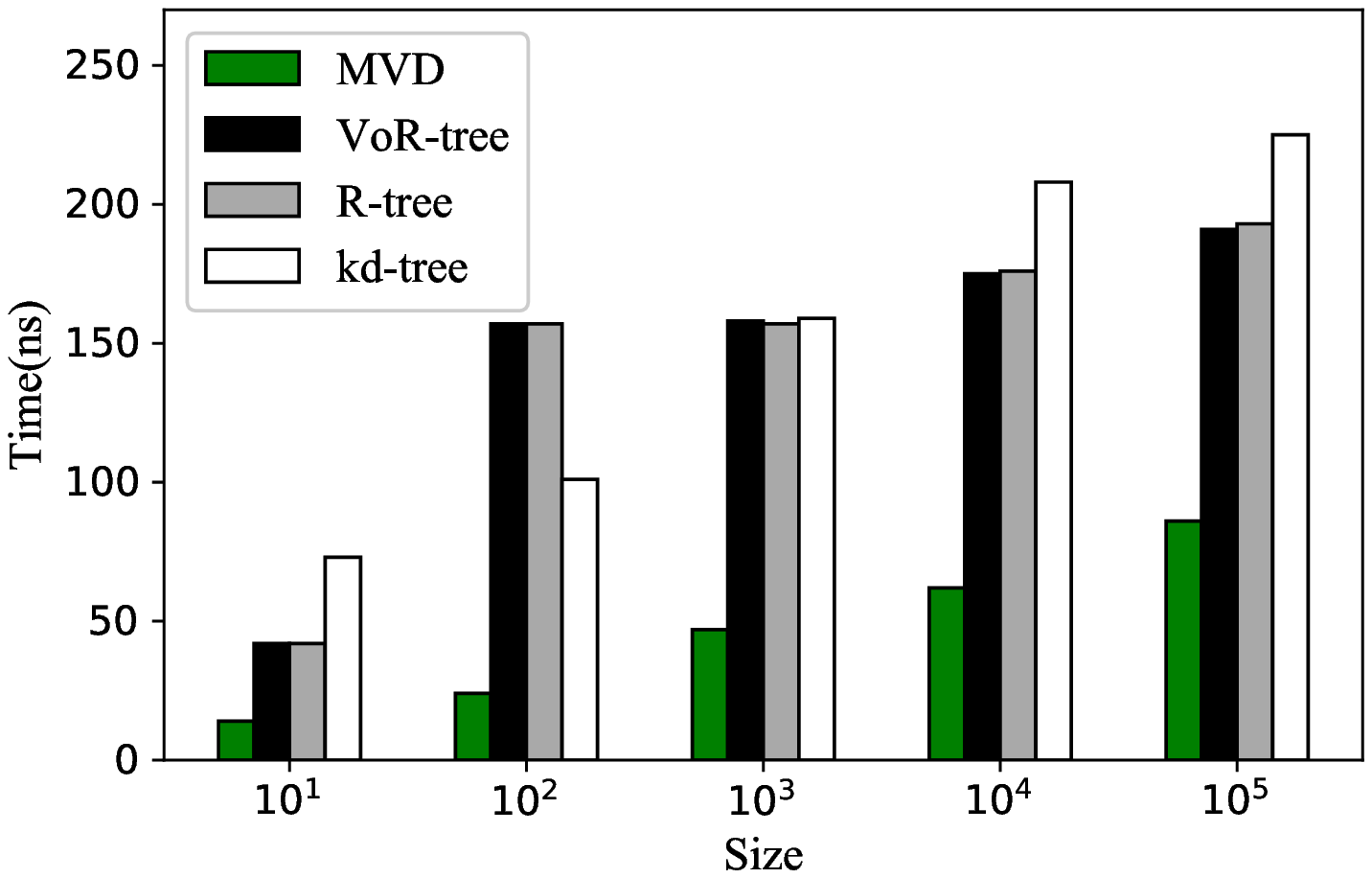}
    \caption{NN from uniform data sets with various sizes}
    \label{nn_uniform}
\end{figure}

\begin{figure}[htbp]
    \centering
    \includegraphics[width=0.48\textwidth]{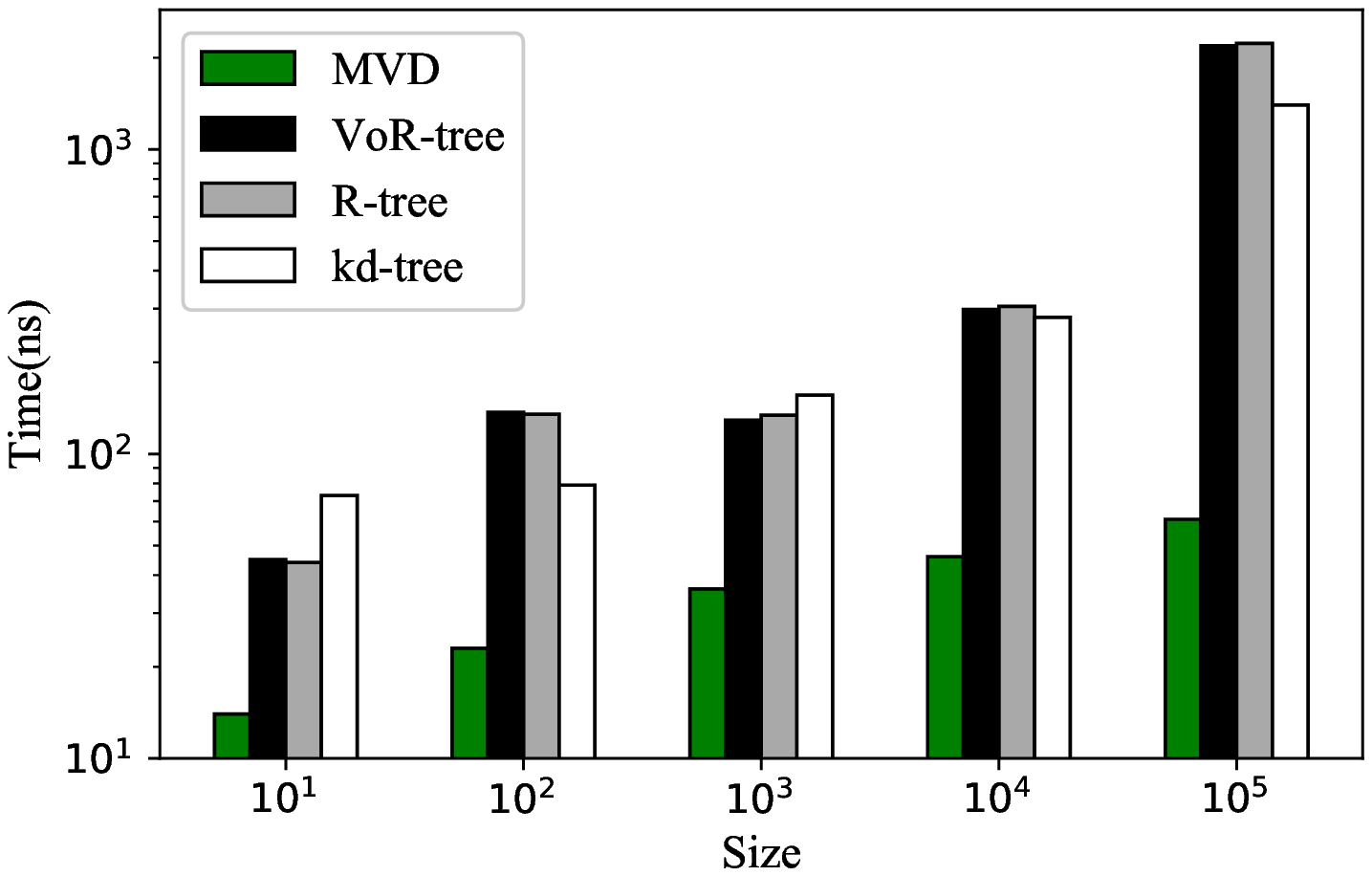}
    \caption{NN from nonuniform data sets with various sizes}
    \label{nn_nonuniform}
\end{figure}

\begin{table*}[htbp]
\centering
\caption{Total computation time (in ns) of NN queries from data sets with various sizes.}
\label{NN_TABLE}
\begin{tabular}{@{}crrrrrrrr@{}}
\toprule
\multirow{2}{*}{\textbf{Size}} & \multicolumn{4}{c}{\textbf{Uniform}} & \multicolumn{4}{c}{\textbf{Nonuniform}} \\ \cmidrule(lr){2-5} \cmidrule(l){6-9}
& \textbf{\quad MVD} & \textbf{\quad VoR-tree}  & \textbf{\quad R-tree}   & \textbf{\quad kd-tree}   & \textbf{\quad MVD} & \textbf{\quad VoR-tree}     & \textbf{\quad R-tree}    & \textbf{\quad kd-tree}    \\ \midrule
$10^1$  & 14      & 42       & 42      & 73      & 14      & 45         & 44       & 72      \\
$10^2$  & 24      & 157      & 157     & 101     & 23      & 137        & 135      & 79     \\
$10^3$  & 47      & 158      & 156     & 159     & 36      & 129        & 134      & 156     \\
$10^4$  & 62      & 175      & 176     & 208     & 46      & 298        & 305      & 281    \\
$10^5$  & 86      & 191      & 193     & 225     & 61      & 2190       & 2231     & 1399   \\ \bottomrule
\end{tabular}
\end{table*}

\begin{figure}[htbp]
    \centering
    \includegraphics[width=0.48\textwidth]{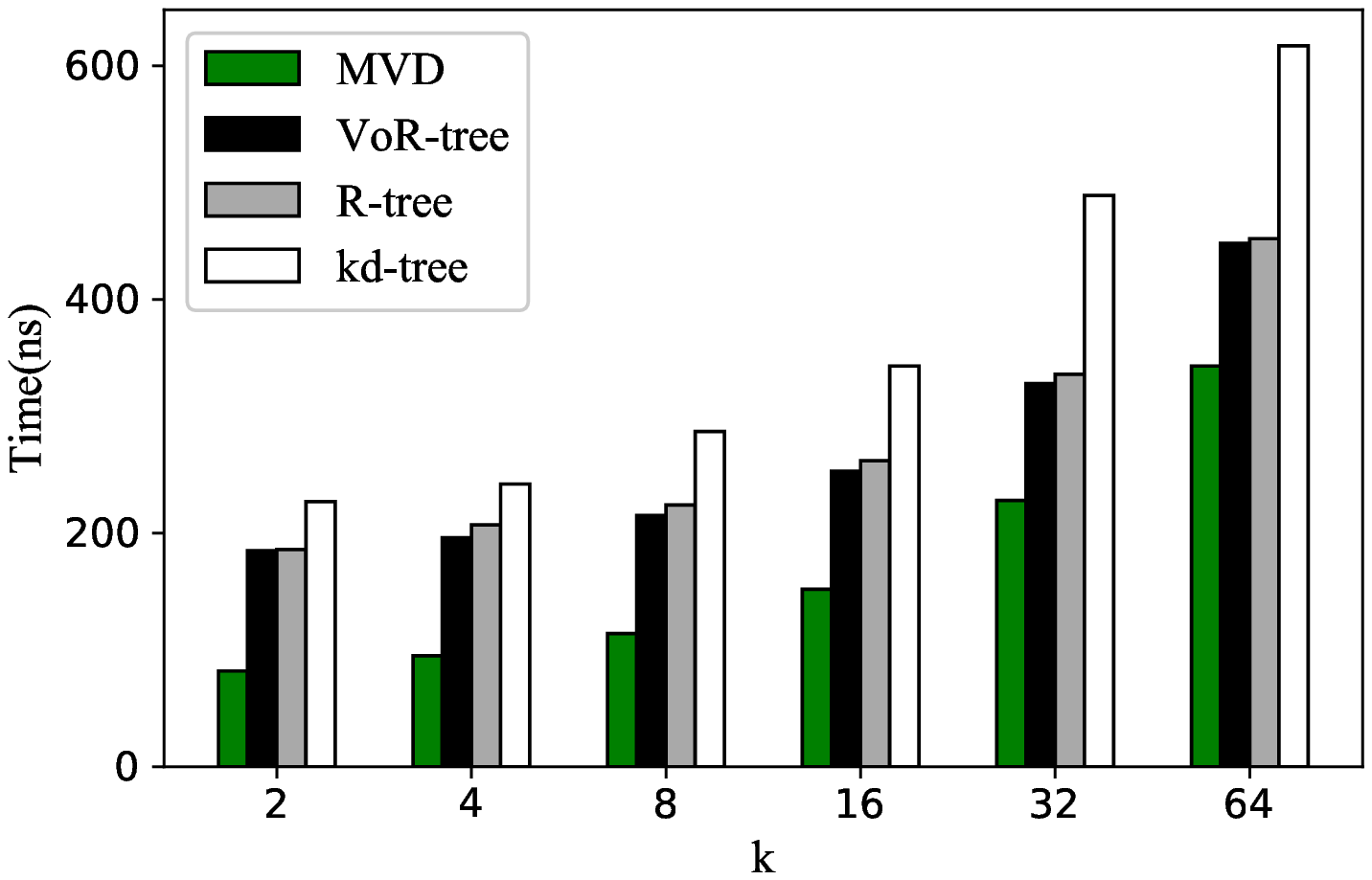}
    \caption{$k$NN from uniform data sets with various $k$}
    \label{knn_uniform}
\end{figure}
\begin{figure}[htbp]
    \centering
    \includegraphics[width=0.48\textwidth]{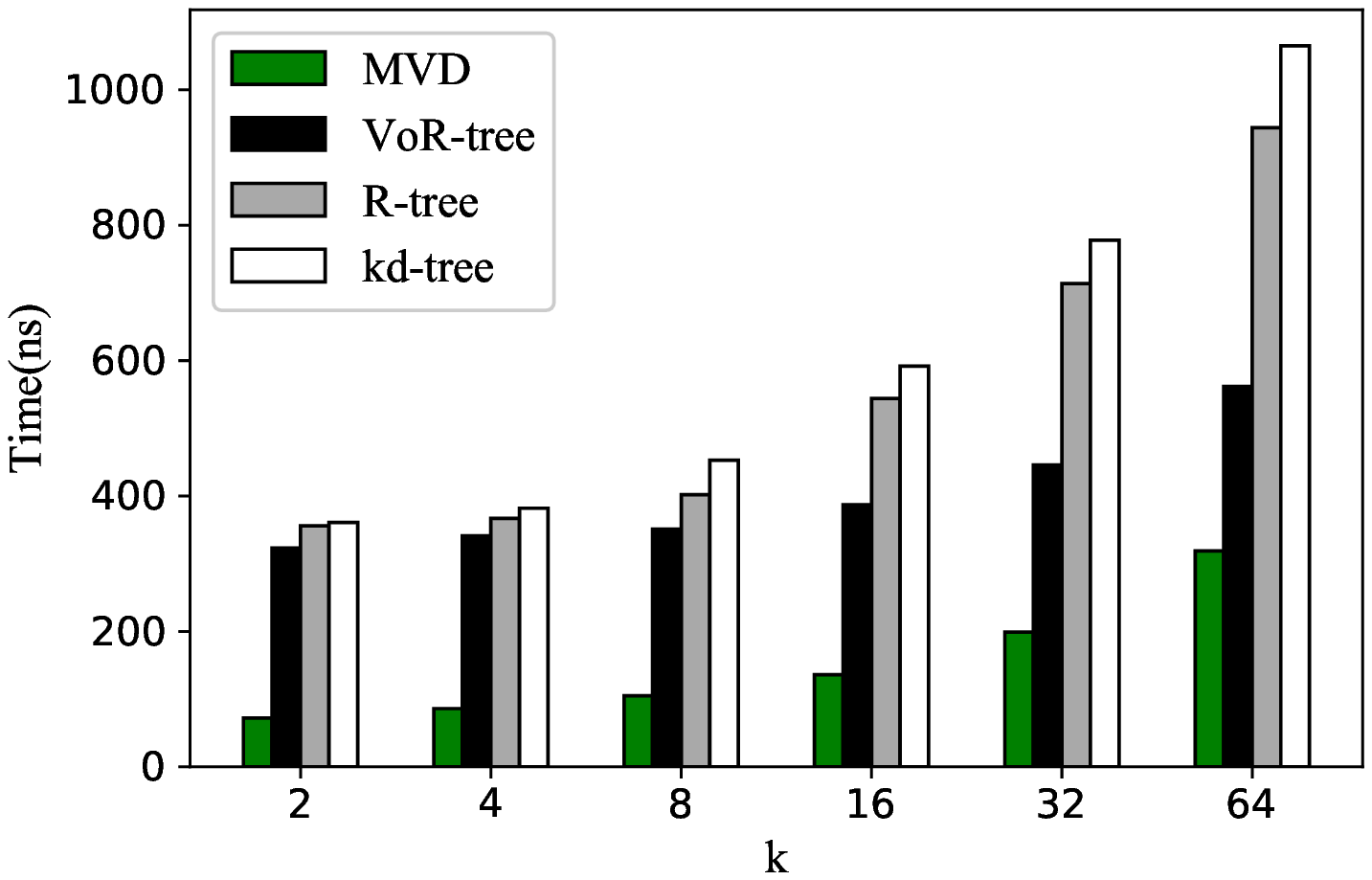}
    \caption{$k$NN from nonuniform data sets with various $k$}
    \label{knn_nonuniform}
\end{figure}
\begin{figure}[htbp]
    \centering
    \includegraphics[width=0.48\textwidth]{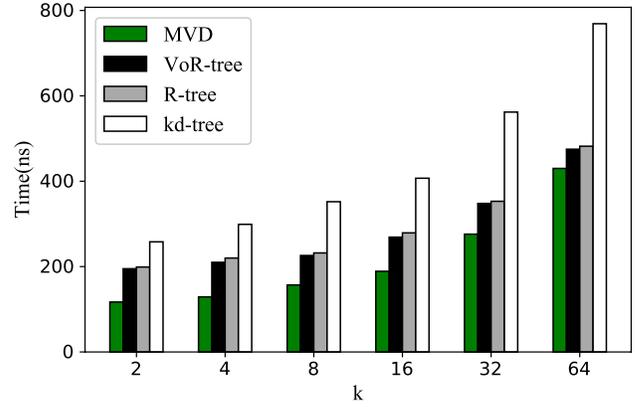}
    \caption{$k$NN from US data set}
    \label{knn_us}
\end{figure}
\begin{table*}[htbp]
\centering
\caption{Total computation time (in ns) of $k$NN queries with various values of $k$.}
\label{KNN_TABLE}
\begin{tabular}{@{}crrrrrrrrrrrr@{}}
\toprule
\multirow{2}{*}{\textbf{$k$}} & \multicolumn{4}{c}{\textbf{Uniform}} & \multicolumn{4}{c}{\textbf{Nonuniform}} & \multicolumn{4}{c}{\textbf{US}} \\ \cmidrule(lr){2-5} \cmidrule(l){6-9} \cmidrule(l){10-13}
& \textbf{\; MVD} & \textbf{\; VoR-tree}  & \textbf{\; R-tree}   & \textbf{\; kd-tree}   & \textbf{\; MVD} & \textbf{\; VoR-tree}     & \textbf{\; R-tree}    & \textbf{\; kd-tree} & \textbf{\; MVD} & \textbf{\; VoR-tree}  & \textbf{\; R-tree}   & \textbf{\; kd-tree}    \\ \midrule
2     & 82       & 185      & 186     & 227      & 72       & 323       & 356      & 361      & 117       & 195       & 199      & 258  \\
4     & 95      & 196      & 207     & 242      & 86       & 341       & 367      & 382      & 129       & 210      & 220      & 299  \\
8     & 114      & 215      & 224     & 287      & 105      & 351       & 402      & 453      & 157       & 226       & 232      & 352  \\
16    & 152      & 253      & 262     & 343      & 136      & 387       & 544      & 592      & 189       & 269       & 279      & 407  \\
32    & 228      & 328      & 336     & 489      & 199      & 446       & 714      & 778      & 276       & 348       & 353      & 562  \\
64    & 343      & 448      & 452     & 617     & 319      & 562       & 944      & 1065     & 430       & 475       & 482      & 769 \\ \bottomrule
\end{tabular} 
\end{table*}

\begin{figure}[htbp]
    \centering
    \includegraphics[width=0.48\textwidth]{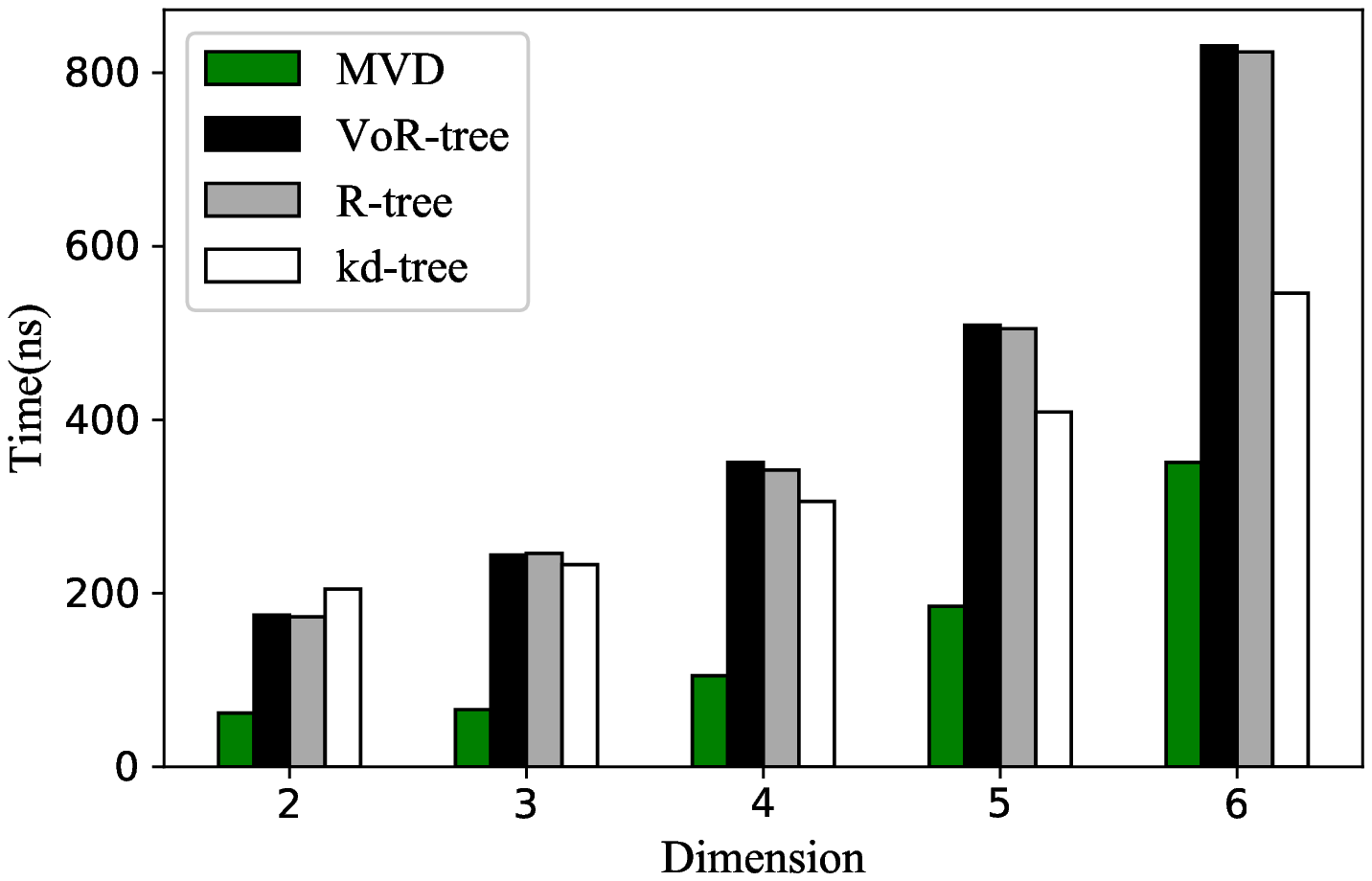}
    \caption{NN from data sets with various dimensions}
    \label{nn_dimension}
\end{figure}

\begin{figure}[htbp]
    \centering
    \includegraphics[width=0.48\textwidth]{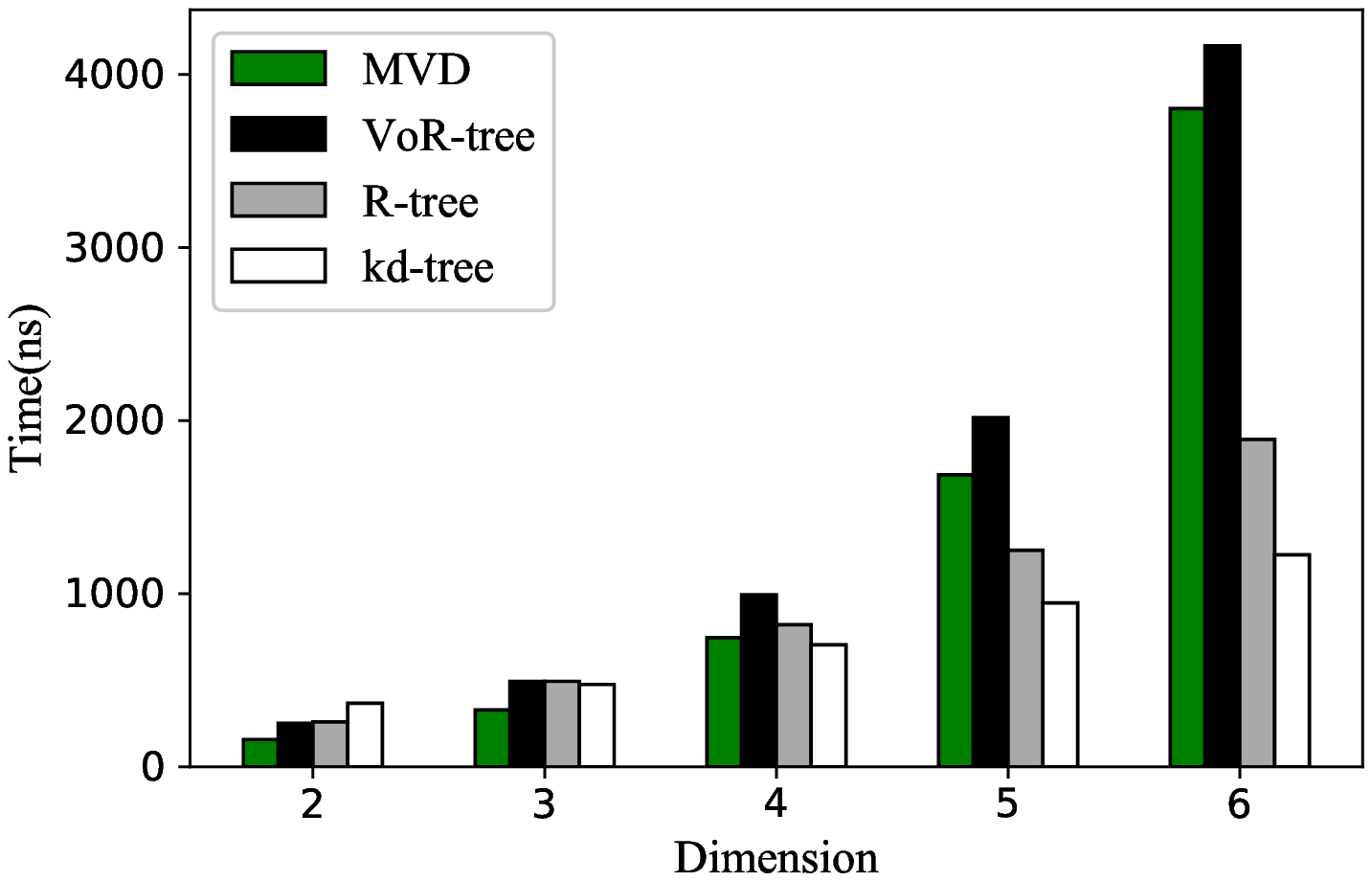}
    \caption{$k$NN from data sets with various dimensions}
    \label{knn_dimension}
\end{figure}

\begin{table*}[htbp]
\centering
\caption{Total computation time (in ns) of NN and $k$NN queries from data sets with various dimensions.}
\label{DIMENSION_TABLE}
\begin{tabular}{@{}crrrrrrrr@{}}
\toprule
\multirow{2}{*}{\textbf{Dimension}} & \multicolumn{4}{c}{\textbf{NN}} & \multicolumn{4}{c}{\textbf{$k$NN}} \\ \cmidrule(lr){2-5} \cmidrule(l){6-9}
& \textbf{\quad MVD} & \textbf{\quad VoR-tree}  & \textbf{\quad R-tree}   & \textbf{\quad kd-tree}   & \textbf{\quad MVD} & \textbf{\quad VoR-tree}     & \textbf{\quad R-tree}    & \textbf{\quad kd-tree}    \\ \midrule
2     & 62       & 175      & 173      & 205       & 158         & 252      & 259       & 368      \\
3     & 66      & 244      & 246      & 233       & 329         & 493      & 494       & 476     \\
4     & 105      & 351      & 342      & 306       & 746         & 994      & 821       & 705     \\
5     & 185      & 509      & 505      & 409       & 1687        & 2018     & 1251      & 947    \\
6     & 351      & 831      & 824      & 546       & 3803        & 4165      & 1891      & 1225    \\\bottomrule
\end{tabular}
\end{table*}

% 在实验中，我们将最常用的两种空间索引结构kd-tree、R-tree，以及最先进的R-tree衍生索引VoR-tree作为对比对象，分别从NN查询和$k$NN查询两个方面验证MVD在空间最近邻上的性能表现。上述的4种索引及其相关算法都是通过Python语言实现的，在参数设置方面，kd-tree的leafSize=100,R-tree和VoR-tree的nodeCapacity=100,MVD的k=100。
In the experiments, we apply the two most common spatial indexing structures kd-tree and R-tree, and the state-of-the-art indexing as a derivative of R-tree: VoR-tree, to be the benchmarks. Thus, we can investigate the performance of MVD on the spatial nearest neighborhood in terms of the NN search and the $k$NN search. These four indexing methods and their related algorithms are implemented using Python. As for the parameter settings, the kd-tree has the leaf size as $100$, R-tree and VoR-tree have the node capacity as 100, and MVD has $k=100$.
% 我们所使用的实验数据有三种形式：模拟生成的随机均匀分布的离散点数据(uniform data)，模拟生成的指数随机分布的离散点数据（简称nonuniform data），现实世界的数据（49603个National Register of Historic Places不重复的地理坐标数据，简称 US data），如图\ref{benchmark}。
There are three types of our datasets used in the experiments: the simulated evenly distributed discrete data, i.e., uniform data, the simulated exponential distributed discrete data, i.e., nonuniform data, and the real-world data which are 49,603 non-repeated geographical coordinate data points from National Register of Historic Places, i.e., the US data. These  data are demonstrated as in Figure~\ref{benchmark}.

% 我们实验的运行环境是这样的：a personal computer with the python2.7, the CPU as Intel Core i5-4308U 2.80GHz and the RAM as DDR3 8G. 
% 因为本文对空间索引的研究都是在内存环境下进行讨论，所以影响性能的主要是算法的CPU总运算量,而CPU的总运算量直接反映就是算法执行的总耗时。所以对于每一种查询方式，我们都是通过对比它们完成一次该查询的时间开销，来评价这些算法性能表现。为了降低实验误差，每项实验我们都进行了30次，并求得时间结果的平均值。
The settings of our experiment environment are as follows. The experiment is conducted on a personal computer with Python 2.7. The CPU is Intel Core i5-4308U 2.80GHz and the RAM is DDR3 8G. As our spatial indexing analysis is in the internal memory environment, what impacts the performance is mainly the total CPU computation of the algorithm. However, the total CPU computation can directly reflect the total time consumed for the execution of the algorithm. Therefore, for every spatial query method, we evaluate the performance of these algorithms based on the comparison of their time cost for one search. To decrease the error of the experiments, we repeat each experiment for $30 \times 1000$ times and calculate the average time cost.
\subsection{Experimental results}
% 对于NN查询，我们通过uniform data和nonuniform data这两种不同分布状态的模拟二维点数据，验证这些索引在不同数据规模下的表现，它们的性能对比如图6和图7所示，其中详细的结果记录见表1。
For the NN search, we use the uniform data and nonuniform data simulated in 2 dimensions from two distributions, in order to evaluate the performance of these indexing methods in different scales of data. The comparison of their performance is presented in Figure~\ref{nn_uniform} and~\ref{nn_nonuniform}. The detailed results are demonstrated in Table~\ref{NN_TABLE}.

% 对于$k$NN查询，我们通过一个uniform data集合 、一个nonuniform data集合和一个US dataset来完成来验证，其中uniform dataset 和nonuniform data的大小都是10000，且维度都是2。它们的性能对比如图8和图9所示，其中详细的结果记录见表2。
For the $k$NN search, we investigate its performance with a set of uniform data, a set of nonuniform data and the US dataset, where the uniform dataset and the nonuniform dataset have the size as 10,000 and 2 dimensions. The comparison of their performance is demonstrated as in Figure~\ref{knn_uniform}, \ref{knn_nonuniform} and \ref{knn_us}. The details of it are presented in Table \ref{KNN_TABLE}.

% 我了验证这些索引性能对维度的敏感性，我们也通过5组不同维度数据集分别验证它们在NN和$k$NN上的性能表现，其中这5组数据集的大小都是10000，且都是采用均匀分布。
To investigate the sensitivity of these indexing methods to the dimension, we evaluate their performance on NN and $k$NN on 5 datasets with different dimensions. The size of these two datasets are all 10,000 and they are all evenly distributed. The comparison of their performance is demonstrated as in Figure~\ref{nn_dimension} and \ref{knn_dimension}. The details of it are presented in Table \ref{DIMENSION_TABLE}.

% 从上述的实验结果，我们可以看出，在2维的情况下，MVD无论在NN还是$k$NN上都优于其他的3种索引结构，尤其在非均匀分布的情况下，其优势更加显著。这都得益于其非树形的索引结构。因为没有树形结构的空间划分，所以在MVD中并不存在节点的从属关系，只有连接关系，任何数据节点也就不会固定从属于某个上级节点。所以我们的索引结构不会出现因失衡而导致的层数太多的情况。在面对NN和$k$NN查询的时候也不会因为节点之间重叠而需要验证过多的节点数据。
From the experiment results above, we can observe that MVD outperforms the other 3 indexing structures no matter in NN or $k$NN for the 2 dimensional case, especially in the case of the uneven distribution. The reason is that MVD has the non-tree indexing structure. Since there is no spatial partition from the tree structure, there is no belonging relations among the nodes in MVD but only the connection relations. No node statically belongs to a certain node in an upper layer. Thus, our indexing structure does not incur the situation where there are too many layers from losing the balance. For the NN and $k$NN search, it is not required to evaluate too many node data from the overlap among nodes.
% 从维度敏感度的测试中，我们可以看出，我们的索引结构与其他的三种索引结构一样，也没有逃出curse of dimensionality\cite{DBLP:conf/vldb/WeberSB98}。随着维度的增加，这4个索引的效率逐渐降低。值得庆幸的是，在NN查询上，随着维度增加，MVD的时间开销已然是低于其他结构的。但是在$k$NN上，MVD和VoR-tree的查询效率在维度较高时（>4）明显低于R-tree和kd-tree。由性质7和性质11可知，高维的维诺图拓扑结构非常复杂；随着维度增加，维诺图中每个节点平均的维诺邻居数量显著增加。我们通过10000个离散点实验列出了2-6维情况下每个点的平均维诺邻居数量（见表4）。从表中可以看出点均维诺邻居量随着维度增加是成指数增长的。MVD和VoR-tree的$k$NN查询都是通过一个维诺图来扩展NN查询而实现的。在高维的情况下，这种增量算法需要验证更多的点才能最终确定的k个最近邻的点，所以效率明显降低。

\begin{table}[htbp]
\centering
\caption{Number of Voroinoi neighbors per generator point}
\label{NVN}
\begin{tabular}{@{}cccccc@{}}
\toprule
\textbf{Dimension} & \textbf{2} & \textbf{3} & \textbf{4} & \textbf{5} & \textbf{6} \\ \midrule
\textbf{Value} & \multicolumn{1}{r}{5.9942} & \multicolumn{1}{r}{15.3938} & \multicolumn{1}{r}{35.6104} & \multicolumn{1}{r}{76.7206} & \multicolumn{1}{r}{153.8450} \\ \bottomrule
\end{tabular}
\end{table}
In the sensitivity test to the dimensions, it can be viewed that our indexing structure is the same as the other indexing structures: the curse of dimensionality \cite{DBLP:conf/vldb/WeberSB98} cannot be avoided. As the dimension increases, the efficiency of these four indexing methods gradually decays. Fortunately, for the NN search, as the dimension becomes larger, the time cost is already less than the other structures. Besides, for $k$NN, the search efficiency of MVD and VoR-tree is evidently less than R-tree and kd-tree in the case when the dimension is larger than 4. From Property 7 and 11, the topological structure of the high-dimiensional Voronoi diagram. As the dimension increases, the average number of Voronoi neighbors significantly increases. We use the experiment of 10,000 discrete points to list the average number of Voronoi neighbors for every point for the dimensions between 2 to 6 as in Table~\ref{NVN}. From this table, the number of the point-wise average number of Voronoi neighbors has the logarithm increase. The $k$NN search methods of MVD and VoR-tree are both through a Voronoi diagram to extend the NN search. For the high-dimensional cases, this incremental algorithm requires to evaluate more points, in order to ensure the nearest k neighbors, so the efficiency is significantly jeopardized.
%这种增量算法需要验证更多的点才能最终确定k个最近邻的点，所以效率明显降低

\section{Conclusions}
% 我们综述了主流的用于NN/$k$-NN查询的结构及方法，分析了传统树形索引在最近邻相关的查询上的不足之处。然后创新性的提出了一种有多层维诺图构成的空间索引结构，MVD。该索引彻底摒弃了树形结构，利用维诺图的邻近探索（neighborhood exploration）能力和多层结构的逐层逼近能力，实现了稳定且高效的NN查询和$k$-NN查询。我们的理论分析和实验结果都表明，在面对现实物理空间的最近邻查询时，MVD的效率是显著高于kd-tree、R-tree和VoR-tree等树形索引的。
We review the major methods for the NN/$k$NN search structures and methods, and analyze the weaknesses of the traditional tree indexing methods in the nearest neighbor search. Then we propose a spatial indexing structure with the Multi-layer   Voronoi diagram. This indexing method abandons the tree structure, uses the neighborhood exploration and the layer-by-layer approaching abilities of the Voronoi diagram, and realizes the stable and efficient NN search and $k$NN search. The theoretical analysis and the experiments indicate that the efficiency of MVD is significantly higher than the tree structure indexing such as kd-tree, R-tree and VoR-tree, in terms of the real physical spatial nearest neighbor search. 

%当然我们也发现MVD也存在着它的局限性：
However, MVD also has its limitations:
\begin{itemize}
    \item The number of realized spatial search methods based on this structure is still quite small. Nevertheless, we already begin the research for the other spatial search method based on this structure, where the range search has achieved the initial success.
    \item Currently, we only propose the MVD construction and nearest neighbor search methods based on the internal memory environment, and are not able to adapt to the extremely large-scale data management circumstance. Therefore, we also started the strategies on the external memory and the distributed environment.
    \item Because of the nature of the Voronoi diagram, MVD is very sensitive to the dimensions. Thus, when the scale of the dataset is not large enough and the dimension is high, its performance is not satisfactory enough. There is still a research gap on how to improve the search efficiency under the MVD structure in the high-dimensional circumstance.
\end{itemize}
% 1、基该结构已经实现的空间查询方式较少，但是我们已经开始了基于该结构的其他空间查询的研究，其中范围查询已经取得了初步成果。

% 2、暂时我们只提出了基于内存环境的MVD构建和空间最近邻查询方法，还无法适应超大规模的数据管理场景，所以我们已经开始着手基于外存和分布式环境的实现策略。
% 3、由于Voronoi diagram的本身特点，MVD对维度非常敏感。所以在数据规模还不足够大，而数据维度又较高的情况下，它的表现并不好。如何在高维度环境下改善基于MVD结构的查询效率还有待研究。
\bibliographystyle{unsrt}
\bibliography{icde2020}
\end{document}